\documentclass{article} 
\usepackage{iclr2025_conference,times}

\usepackage{amsmath,amsfonts,bm}

\def\eqref#1{equation~\ref{#1}}

\def\1{\bm{1}}

\DeclareMathAlphabet{\mathsfit}{\encodingdefault}{\sfdefault}{m}{sl}
\SetMathAlphabet{\mathsfit}{bold}{\encodingdefault}{\sfdefault}{bx}{n}

\usepackage{hyperref}
\usepackage{url}
\usepackage{amssymb}
\usepackage{graphicx}
\usepackage{booktabs}

\title{SWE-TRACE: Optimizing Long-Horizon SWE Agents through Rubric Process Reward Models and Heuristic Test-Time Scaling}

\author{Hao Han\thanks{These authors contributed equally to this work.}, Jin Xie\footnotemark[1], Xuehao Ma\footnotemark[1], Weiquan Zhu, Ziyao Zhang, \\
\textbf{ZhiLiang Long, Hongkai Chen\thanks{ Corresponding author, allenhkchen@gmail.com.} and Qingwen Ye} \\
vivo, ShenZhen, China \\
}

\iclrfinalcopy 
\begin{document}

\maketitle

\begin{abstract}
Resolving real-world software engineering (SWE) issues with autonomous agents requires complex, long-horizon reasoning. Current pipelines are bottlenecked by unoptimized demonstration data, sparse execution rewards, and computationally prohibitive inference scaling, which collectively exacerbate token bloat, reward hacking, and policy degradation. We present SWE-TRACE (Trajectory Reduction and Agentic Criteria Evaluation), a unified framework optimizing the SWE agent lifecycle across data curation, reinforcement learning (RL), and test-time inference. First, we introduce an LLM multi-task cascading method, utilizing step-wise oracle verification to distill a 60K-instance Supervised Fine-Tuning (SFT) corpus strictly biased toward token-efficient, shortest-path trajectories. Second, to overcome the instability of sparse outcome rewards, we design a Memory-Augmented Agentic RL pipeline featuring a Rubric-Based Process Reward Model (PRM). An auxiliary Rubric-Agent provides dense, fine-grained heuristic feedback on intermediate steps, guiding the model through long-horizon tasks. Finally, we bridge training and inference by repurposing the PRM for heuristic-guided Test-Time Scaling (TTS). By dynamically evaluating and pruning action candidates at each step, SWE-TRACE achieves superior search efficiency without the latency overhead of standard parallel sampling. Extensive experiments on standard SWE benchmarks demonstrate that SWE-TRACE significantly advances the state-of-the-art, maximizing resolution rates while drastically reducing both token consumption and inference latency.
\end{abstract}

\section{Introduction}
Large language models (LLMs) are rapidly evolving from passive code generators into autonomous software engineering (SWE) agents that can read issue reports, navigate repositories, edit files, run tests, and iteratively refine patches in realistic development environments~\cite{ wang2025openhandssoftwareagentsdk, yang2024sweagentagentcomputerinterfacesenable}. This shift has been catalyzed by the emergence of repository-level benchmarks such as SWE-bench~\cite{jimenez2024swebenchlanguagemodelsresolve}, which reframed software engineering as an end-to-end problem grounded in real GitHub issues rather than isolated function completion. At the same time, systems such as SWE-agent and OpenHands have demonstrated that tool-using, ReAct-style~\cite{yao2023reactsynergizingreasoningacting} interaction loops are substantially more effective for repository-scale tasks than one-shot code generation, because successful issue resolution typically requires multi-step exploration, debugging, execution, and revision over long interaction horizons.

Despite rapid progress, building strong open SWE agents remains difficult for three reasons. First, supervised trajectories are often inefficient: many agent traces contain redundant exploration, repeated tool use, and unnecessarily long reasoning chains, so fine-tuning on them teaches the model to imitate noisy search instead of efficient problem solving~\cite{yang2025swesmithscalingdatasoftware, jain2025r2egymproceduralenvironmentshybrid}. Second, reinforcement learning for SWE is a long-horizon credit assignment problem: outcome rewards are typically sparse and delayed, since a trajectory may contain many steps while success is judged only by final execution results~\cite{shum2025swermexecutionfreefeedbacksoftware, luo2025deepswe}. This makes it hard to identify which intermediate actions were useful and can encourage inflated or unstable behavior. Third, test-time scaling (TTS)~\cite{muennighoff2025s1simpletesttimescaling, zhang2025surveytesttimescalinglarge} is expensive: existing gains often rely on sampling many full trajectories and reranking them, which introduces substantial latency in repository-level environments. Recent work on large-scale SWE data, RL-based SWE agents, and execution-free reward models has made strong progress on each of these axes~\cite{shum2025swermexecutionfreefeedbacksoftware}, but these challenges remain only partially addressed when data curation, RL, and inference-time search are optimized separately.

In this paper, we present SWE-TRACE, a unified framework for training and inference of long-horizon SWE agents. Our key idea is to optimize the entire agent pipeline around process efficiency. At the supervised stage, the model should learn from trajectories that reflect short, high-fidelity solution paths rather than verbose exploration. At the reinforcement learning stage, optimization should be guided not only by final execution outcomes, but also by dense intermediate process signals. At inference time, verification should be used not only to rerank completed trajectories, but also to steer action selection early enough to avoid wasting computation on weak branches.

Based on this principle, SWE-TRACE introduces a three-stage pipeline. First, we construct a massive-scale SWE training corpus by scaling synthetic issue generation across 77 repositories and filtering 140K candidate instances into 60K high-quality samples, together with distilled trajectories from both frontier closed-source and strong open-source teachers. Second, we propose token-efficient trajectory optimization through LLM multi-task cascading, where multiple candidate actions are generated at each step and an oracle verifier selects the best continuation, producing shorter and cleaner supervision traces. Third, we develop memory-augmented agentic reinforcement learning with rubric-based process reward models, and further reuse the learned rubric signals for heuristic-guided low-latency test-time scaling that prunes poor actions during rollout instead of only reranking complete trajectories afterward.

Our experiments show that this integrated recipe substantially improves the SWE capabilities of lightweight 4B and 30B models on SWE-bench Verified, narrowing the gap between compact open models and much larger frontier systems. More broadly, our results suggest that the path toward stronger SWE agents is not simply to scale model size or rollout count, but to improve how agents are taught, rewarded, and guided throughout the full lifecycle of decision making.

In summary, our main contributions are as follows:
\begin{itemize}
    \item We present a massive-scale SWE data curation pipeline that expands synthetic issue construction to 77 repositories and produces 60K high-quality training instances from 140K candidates.
    \item We propose LLM multi-task cascading for token-efficient trajectory synthesis, yielding shorter and higher-fidelity SFT supervision by selecting strong stepwise continuations with oracle verification.
    \item We introduce memory-augmented agentic RL with rubric-based process reward models, enabling dense, interpretable, and design-aware feedback beyond sparse execution rewards.
    \item We develop a heuristic-guided low-latency test-time scaling method that reuses the trained PRM to guide step-level action sampling during inference.
    \item We demonstrate that the resulting framework elicits strong long-horizon SWE capability in 4B and 30B models, achieving competitive performance on SWE-bench Verified under practical compute budgets.
\end{itemize}

\section{Related Work}
\label{Related Work}
\subsection{Software engineering benchmarks and agent frameworks}
Recent progress in software engineering agents has been driven by repository-level benchmarks and increasingly capable agent scaffolds. SWE-bench~\cite{jimenez2024swebenchlanguagemodelsresolve} established real-world GitHub issue resolution as a challenging benchmark for language models, and SWE-bench Verified later introduced a human-validated subset of 500 instances to improve evaluation reliability. On the systems side, SWE-agent~\cite{yang2024sweagentagentcomputerinterfacesenable} showed that specialized agent-computer interfaces substantially improve repository navigation, editing, and execution, while OpenHands~\cite{wang2025openhandssoftwareagentsdk} generalized this paradigm into an extensible platform for tool-using software agents. In parallel, Agentless~\cite{xia2024agentlessdemystifyingllmbasedsoftware} demonstrated that competitive performance can also arise from simpler localization–repair–validation pipelines, highlighting that the design space includes both full agentic interaction loops and more structured non-agentic decomposition.

\subsection{Scalable data construction and executable training environments}
A major recent trend is to move from evaluation-only benchmarks toward scalable training environments and synthetic data pipelines. SWE-Gym~\cite{pan2025trainingsoftwareengineeringagents} introduced one of the first open training environments for SWE agents, with 2,438 executable tasks drawn from real repositories. More recently, SWE-smith~\cite{yang2025swesmithscalingdatasoftware} proposed a scalable data-generation pipeline that synthesizes large numbers of bug-fixing tasks directly from codebases, producing 50K instances from 128 repositories. R2E-Gym~\cite{jain2025r2egymproceduralenvironmentshybrid} pushed this line further with a procedurally curated executable environment of over 8K tasks and a detailed study of verifier-guided test-time scaling. These works collectively show that data scale and environment availability are becoming first-class bottlenecks in SWE-agent research.

\subsection{Reinforcement learning, reward models, and process supervision}
Another active direction focuses on optimizing coding agents with reinforcement learning or learned verifiers. Earlier work such as CodeRL~\cite{le2022coderlmasteringcodegeneration} showed that critic-style feedback can improve program synthesis beyond standard supervised learning. In the SWE setting, recent works such as DeepSWE~\cite{luo2025deepswe} and SWE-Master~\cite{song2026swemasterunleashingpotentialsoftware} show that long-horizon SWE ability can be substantially improved through post-training with execution environments, large-scale rollouts, and test-time scaling. At the same time, SWE-RM~\cite{shum2025swermexecutionfreefeedbacksoftware} argues that verifier quality for SWE cannot be judged by test-time scaling alone, and that discrimination and calibration are also important if a verifier is to serve as a stable RL signal. More broadly, this connects to the growing literature on process supervision, where step-level feedback has been shown to be more informative than pure outcome supervision in complex reasoning tasks.

\section{Token-Efficient Trajectory Synthesis.}
\label{sec:SFT}
Recent SWE training environments have rapidly increased the scale of executable data, from SWE-Gym with 2,438 real-world tasks, to SWE-smith with 14K synthetic instances from 114 repositories, and R2E-Gym with more than 4.6K executable tasks and hybrid verifiers. Very recent systems such as Scale-SWE~\cite{zhao2026immersiongithubuniversescaling} and SWE-Next~\cite{liang2026swenextscalablerealworldsoftware} further show that scalable SWE progress increasingly depends on data factories that can construct large numbers of executable, self-verifying task instances. However, raw scale alone is not sufficient for training long-horizon agents. In practice, synthetic bug construction often fails because the perturbation scope is too broad, the target code is weakly connected to executable tests, or the resulting trajectories contain large amounts of redundant exploration. To address these issues, we design a token-efficient synthesis pipeline with two components: (i) a massive-scale, test-grounded data curation framework that constructs high-quality SWE instances from a large repository pool, and (ii) a step-wise trajectory optimization method, LLM multi-task cascading, that compresses successful rollouts into short, high-fidelity SFT traces.

\subsection{Massive-Scale SWE Data Curation and High-Fidelity Trajectory Synthesis}
\textbf{Repository screening and environment onboarding} We begin with a large pool of more than 1,000 GitHub repositories, extending beyond the repositories used in prior synthetic SWE datasets. Since large-scale synthesis is only practical when repositories are executable and testable, we first apply repository-level screening based on two hard constraints: (1) the repository can be built inside Docker, and (2) the repository contains runnable test cases. Repositories that fail to satisfy either condition are discarded early. After this screening stage, we retain 77 repositories for full data construction.

A key challenge in this stage is that many repositories do not come with ready-to-use, stable Docker environments. To make large-scale data construction feasible, we use an agent-based onboarding pipeline, driven primarily by MiniMax 2.5, to automatically build or repair repository Docker environments, resolve missing dependencies, validate test executability, and prepare the repository for downstream synthesis. This agent is also used to help construct data samples once the environment is stabilized. In practice, automating this step is critical: manually containerizing and validating dozens of repositories would otherwise become a major bottleneck.

\textbf{Repository screening and environment onboarding} A central limitation of broad function-level perturbation is that many generated bugs fall outside the executable test surface, making them difficult to validate or unnecessarily hard to synthesize into realistic issue instances. To improve synthesis precision, we introduce a test-grounded scope selection strategy.

For each repository, we first execute its test suite and build a mapping between test cases and repository functions. Let
\[
\mathcal{T} = \{t_1, \dots, t_M\}
\]
denote the set of runnable tests, and let
\[
\mathcal{F} = \{f_1, \dots, f_N\}
\]
denote the set of candidate functions in the repository. After test execution, we construct a relevance mapping
\[
\Gamma : \mathcal{T} \rightarrow 2^{\mathcal{F}},
\]
where $\Gamma(t)$ contains the functions associated with test $t$. In practice, $\Gamma$ is derived from test execution signals together with repository structural information, and can be viewed as a test-to-function relevance graph.

This step defines the synthesis scope \emph{before} bug injection. Rather than allowing the generator to modify arbitrary functions in the repository, we restrict synthesis to functions that are relevant to executable tests. Formally, given a selected subset of tests $\mathcal{T}_{\mathrm{sel}} \subseteq \mathcal{T}$, we define the perturbation scope as
\[
\mathcal{F}_{\mathrm{rel}} = \bigcup_{t \in \mathcal{T}_{\mathrm{sel}}} \Gamma(t).
\]
This substantially narrows the search space and increases the probability that an injected perturbation will produce a meaningful and verifiable bug. Compared with repository-wide synthesis, this test-grounded design yields a much higher success rate because the target region is already known to lie on the tested behavioral surface.

\textbf{Test-aware bug synthesis.}
Once the test-to-function mapping is built, we synthesize bugs only over the scoped function set $\mathcal{F}_{\mathrm{rel}}$. Unlike pipelines that rewrite functions globally, our method performs \emph{test-conditioned synthesis}: the LLM is given not only the target function context, but also the relevant test-case information. This allows the model to reason about expected behavior, identify the semantic contract encoded by the tests, and inject perturbations that break that behavior while remaining plausible.

This design provides two advantages. First, it improves the \emph{success rate of bug synthesis}, because the perturbation target is already anchored to functions that are causally connected to executable tests. Second, it improves the \emph{quality of issue construction}, because the generated issue, failing behavior, and target patch are grounded in the same executable test semantics. In effect, the tests provide a natural semantic anchor for both bug generation and subsequent repair. Table~\ref{tab:test_aware_bug_synthesis} summarizes the effectiveness of test-aware bug synthesis.

\begin{table}[h]
\centering
\caption{Effect of test-aware bug synthesis on benchmark construction success.}
\label{tab:test_aware_bug_synthesis}
\scalebox{0.9}{\begin{tabular}{lcccc}
\toprule
 & \textbf{\# Repo} & \textbf{\# Bug Issue} & \textbf{\# Filtered Sample} & \textbf{Success Rate} \\ \midrule
w/o Test-aware bug synthesis & 25                          & 59002                            & 20638                                  & 35.0\%  \\
\textbf{w. Test-aware bug synthesis} & 25                          & 49318                            & 24995                                  & \textbf{50.7\%} \\ \bottomrule
\end{tabular}}
\end{table}

\textbf{Data construction and filtering.}
Using the above procedure, we generate approximately 140K candidate bug instances across the 77 selected repositories. Each candidate instance consists of an issue description, repository snapshot, target tests, and hidden construction metadata used only during data generation. We then apply rigorous filtering to obtain 60K high-quality samples based on the same agent-based pipeline.

Our filtering protocol has four stages. First, we require \emph{environment validity}: the Dockerized repository must build and run reliably. Second, we require \emph{test validity}: the synthesized bug must induce reproducible fail-to-pass behavior under the chosen tests without collapsing the repository into an unusable state. Third, we require \emph{issue consistency}: the generated issue description must align with the induced failure while avoiding direct leakage of the repair. Fourth, we apply \emph{difficulty and stability filtering}, removing trivial samples that require almost no reasoning as well as unstable samples whose failures are non-deterministic.

\textbf{Hybrid teacher distillation for SFT trajectories.}
On top of the filtered bug corpus, we generate SFT trajectories using a hybrid teacher pool. We combine a strong frontier closed-source teacher (e.g., Claude) with a strong open-source teacher (primarily \texttt{MiniMax 2.5}) to balance trajectory quality, behavioral diversity, and collection cost. In practice, the teacher models are complementary: stronger closed-source models often produce cleaner long-horizon decomposition and more reliable repair behavior, while strong open-source teachers provide broader rollout coverage and more diverse action styles. We retain successful trajectories with executable evidence and use them as the raw supervision source for the next stage.

\subsection{LLM Multi-Task Cascading for Token-Efficient Trajectory Optimization}
\label{subsec:cascading}
Although teacher-generated trajectories can solve many tasks, they are often far from token-efficient. Even successful rollouts may include repeated file inspection, redundant shell commands, exploratory edits that are later discarded, or repeated validation steps with little marginal value. To transform these raw rollouts into better supervision, we introduce \emph{LLM multi-task cascading}, a step-wise optimization procedure that selects high-utility actions while pruning redundant exploration.

\textbf{Multi-task candidate generation.}
At step $t$, let the current interaction history be
\[
h_t = (I, o_0, a_0, \dots, o_{t-1}, a_{t-1}),
\]
where $I$ is the issue description, $a_j$ is an agent action, and $o_j$ is the resulting environment observation. Instead of sampling a single next action, we generate multiple candidate actions under a set of task-specific generation modes:
\[
\mathcal{M} = \{\texttt{localize},\ \texttt{inspect},\ \texttt{edit},\ \texttt{validate},\ \texttt{summarize}\}.
\]
For each mode $m \in \mathcal{M}$, the teacher proposes $K_m$ candidate actions,
\[
\mathcal{A}_t^{(m)} = \{a_{t,1}^{(m)}, \dots, a_{t,K_m}^{(m)}\},
\]
and the total candidate pool is
\[
\mathcal{A}_t = \bigcup_{m \in \mathcal{M}} \mathcal{A}_t^{(m)}.
\]

This differs from ordinary best-of-$N$ decoding. The goal is not merely to obtain multiple stochastic continuations of the same prompt, but to elicit \emph{different operational intents} at each step. For example, one candidate may focus on pinpointing the failure location, another may inspect a relevant call path, and another may attempt a direct code edit. This structured candidate pool increases useful action diversity without losing control over the rollout.

\textbf{Oracle verification with test-grounded supervision.}
Because our instances are synthetically constructed, the data pipeline has access to hidden construction metadata during generation time, including the relevant tests and the target repair region. We use this information only inside a generation-time \emph{oracle verifier} to evaluate candidate actions.

For each candidate action $a \in \mathcal{A}_t$, the oracle executes the action in a sandboxed branch (or a lightweight test mode when possible) and computes a progress score
\begin{equation}
\label{eq:oracle_score}
S(h_t, a)
=
\lambda_1 \Delta_{\mathrm{test}}(h_t, a)
+
\lambda_2 \Delta_{\mathrm{scope}}(h_t, a)
+
\lambda_3 \Delta_{\mathrm{patch}}(h_t, a)
+
\lambda_4 \Delta_{\mathrm{info}}(h_t, a)
-
\lambda_5 C_{\mathrm{tok}}(a)
-
\lambda_6 C_{\mathrm{red}}(h_t, a).
\end{equation}
Here,
\begin{itemize}
    \item $\Delta_{\mathrm{test}}$ measures whether the action improves the current test status;
    \item $\Delta_{\mathrm{scope}}$ measures whether the action moves the trajectory closer to the test-relevant function region;
    \item $\Delta_{\mathrm{patch}}$ measures whether a proposed edit is aligned with the hidden repair direction;
    \item $\Delta_{\mathrm{info}}$ rewards actions that reveal useful debugging information;
    \item $C_{\mathrm{tok}}$ penalizes token-heavy actions;
    \item $C_{\mathrm{red}}$ penalizes redundant behavior, such as repeated file reads, repeated test runs without meaningful edits, or navigation loops.
\end{itemize}

The selected action is
\[
a_t^{*} = \arg\max_{a \in \mathcal{A}_t} S(h_t, a).
\]
The environment is then updated with $a_t^{*}$, and the process repeats until the issue is resolved or the rollout budget is exhausted.

This oracle differs from standard trajectory-level reranking in two ways. First, it operates \emph{step by step}, so weak branches are pruned before they grow into long and expensive trajectories. Second, it is \emph{progress-aware} rather than purely outcome-aware: it can reward actions that measurably improve localization or patch quality even before final resolution.

\textbf{Cascaded shortest-path optimization.}
The cascading procedure can be viewed as greedily approximating the following objective:
\begin{equation}
\label{eq:shortest_path_obj}
\tau^{\dagger}
=
\arg\min_{\tau \in \mathcal{T}(x)}
|\tau|
\quad
\text{s.t.}
\quad
\mathrm{Resolved}(\tau)=1,
\end{equation}
where $|\tau|$ denotes the token or step cost of trajectory $\tau$. The exact shortest successful trajectory is generally intractable, so we approximate it by step-wise oracle selection. In practice, this removes a large fraction of unproductive exploration from successful teacher rollouts.

After a successful cascaded rollout is obtained, we apply a second compression pass to remove low-utility actions whose deletion does not change final executability. This post-processing primarily eliminates: (i) repeated repository inspection after the relevant test-linked region is already identified; (ii) repeated validation commands without intervening material edits; (iii) exploratory edits that are later overwritten; and (iv) verbose shell interaction that does not change the problem-solving state. The result is a shortest-path-style SFT trajectory that preserves the causal structure of debugging while substantially reducing token overhead.

\textbf{Hard negatives from rejected actions.}
An additional benefit of cascading is that rejected candidates become structured hard negatives. At each state, the candidates in
\[
\mathcal{A}_t \setminus \{a_t^{*}\}
\]
are plausible but inferior alternatives under the same context. We retain these rejected actions as state-conditioned negative examples, which later provide useful supervision for downstream process reward modeling and action ranking.

\section{Process-Guided Agentic Reinforcement Learning}
\label{Agentic_RL}
\begin{figure*}[ht]
\begin{center}
\centerline{\includegraphics[width=145mm]{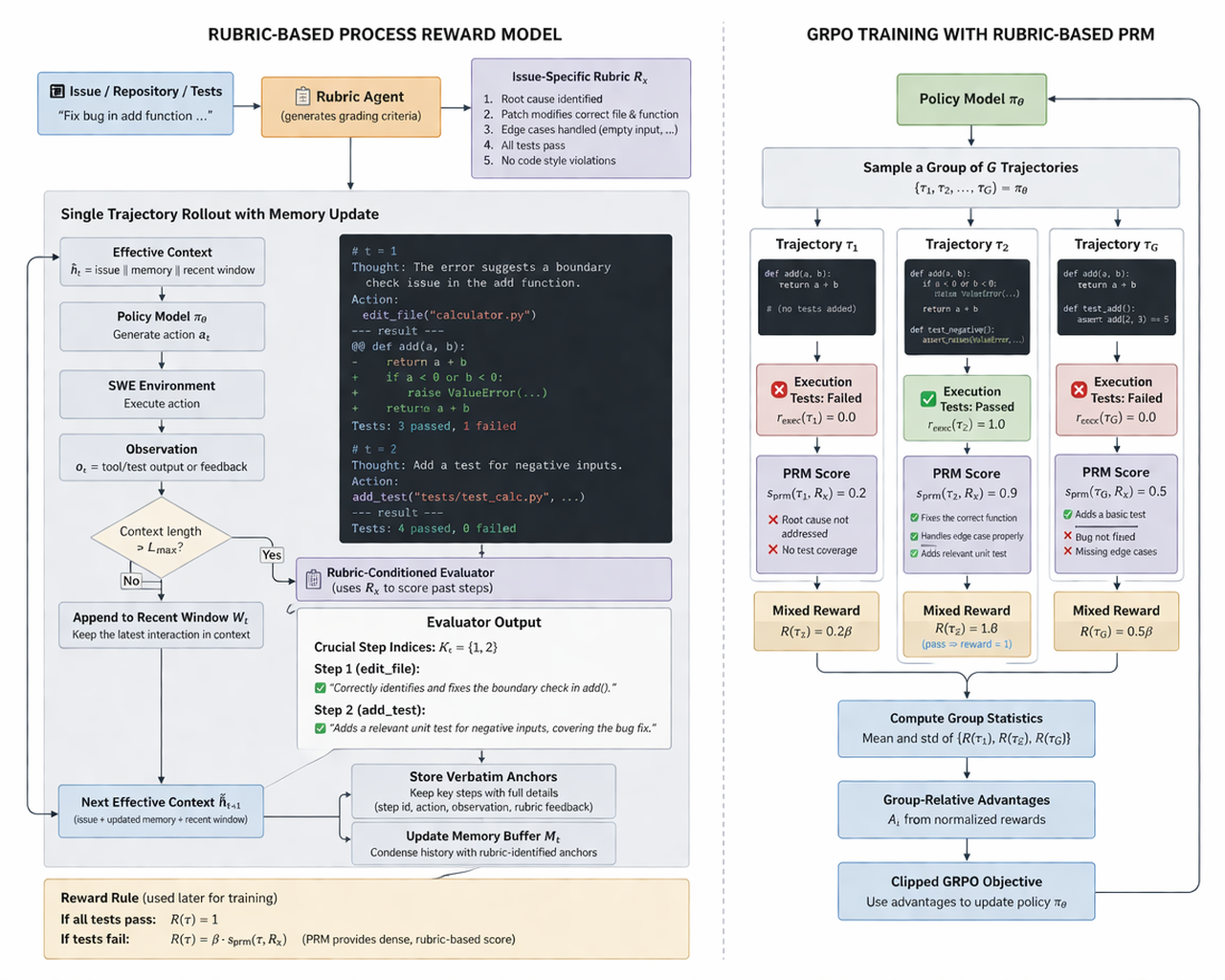}}
\caption{\textbf{Overview of the Rubric-Based PRM and GRPO traning with Rubric-Based PRM.} }
\label{Rubric_PRM_GRPO}
\end{center}
\end{figure*}

Supervised trajectory synthesis provides a strong initialization, but it does not fully solve the optimization problem of long-horizon SWE agents. In realistic repository-scale environments, a trajectory may span tens to hundreds of interaction steps, while the most common reward signal remains terminal and execution-based: the final patch either passes or fails the test suite. Such feedback is sparse, delayed, and often uninformative about \emph{which} intermediate decisions actually contributed to success. Motivated by recent work on process supervision, reward modeling, and long-horizon context management~\cite{lightman2023letsverifystepstep,shao2024deepseekmath,shum2025swermexecutionfreefeedbacksoftware,packer2024memgptllmsoperatingsystems,wang2026sweprunerselfadaptivecontextpruning}, we introduce a \emph{process-guided agentic RL} framework with two coupled components: (i) a \emph{rubric-based process reward model} (PRM) that provides dense and interpretable guidance over intermediate steps, and (ii) a \emph{memory-augmented architecture} that preserves critical evidence when the interaction history exceeds the context budget. Figure~\ref{Rubric_PRM_GRPO} presents the holistic architecture of Rubric-Based PRM and GRPO traning with Rubric-Based PRM.

\subsection{Memory-Augmented Long-Horizon Architecture}
\label{subsec:memory_rl}

Let the agent interact with a SWE environment for an instance $x=(I,C,U)$, where $I$ is the issue, $C$ is the repository state, and $U$ is the test suite. At step $t$, the raw interaction history is
\[
h_t = (I, o_0, a_0, o_1, a_1, \dots, o_{t-1}, a_{t-1}, o_t),
\]
where $a_j$ is an action and $o_j$ is the corresponding environment observation. In long-horizon settings, the token length $\ell(h_t)$ may eventually exceed the model context budget $L_{\max}$, leading to context explosion and degraded reasoning.

To address this, we maintain a \emph{memory buffer} $\mathcal{M}_t$ that stores only \emph{critical steps} from past interaction, while always retaining the most recent short-term window $\mathcal{W}_t$. The effective policy context is
\[
\tilde{h}_t = (I, \mathcal{M}_t, \mathcal{W}_t).
\]
When $\ell(h_t) \leq L_{\max}$, we simply set $\mathcal{M}_t = \varnothing$ and $\mathcal{W}_t = h_t$. When $\ell(h_t) > L_{\max}$, we trigger memory construction and retain only high-value historical evidence.

A key design choice is that memory entries are not abstractive summaries. Instead, we preserve \emph{verbatim anchors} from critical steps:
\[
m_j = \big(j, a_j, o_j\big),
\]
where $j$ is the original step index. This means the agent stores the exact action text, tool outputs, and observations from selected steps, together with their step numbers. We avoid aggressive paraphrasing because long-horizon coding trajectories are highly sensitive to precise details such as filenames, stack traces, command outputs, and patch content; rewriting them can introduce hallucinated or omitted facts. The explicit step index also preserves temporal grounding, allowing the model to recover the original decision chronology.

To decide which historical steps should be retained, we use the trained PRM as a \emph{critical-step detector}. Let $s_t^{\mathrm{prm}}$ denote the PRM score for the current step (defined in Section~\ref{subsec:rubric_prm}). We define the critical-step set as
\[
\mathcal{K}_t
=
\Big\{
j \leq t :
s_j^{\mathrm{prm}} \geq \delta_{\mathrm{abs}}
\ \ \text{or} \ \
|s_j^{\mathrm{prm}} - s_{j-1}^{\mathrm{prm}}| \geq \delta_{\mathrm{chg}}
\Big\},
\]
where $\delta_{\mathrm{abs}}$ selects intrinsically important steps and $\delta_{\mathrm{chg}}$ captures decision points that significantly alter progress. The memory buffer is then
\[
\mathcal{M}_t = \{m_j : j \in \mathcal{K}_t\},
\]
subject to a fixed token budget. If the selected memory still exceeds the budget, we keep the top-scoring entries according to $s_j^{\mathrm{prm}}$ while always preserving the latest local interaction window. This design turns memory into a process-aware retrieval mechanism: the agent does not try to remember everything, but it preserves the exact steps that most strongly determined success or failure.

\subsection{Why Execution-Only Rewards Fail in Long-Horizon SWE}
\label{subsec:sparse_reward}

For a complete trajectory
\[
\tau = (o_0, a_0, o_1, a_1, \dots, o_T, a_T),
\]
the standard execution reward is usually defined as
\[
r_{\mathrm{exec}}(\tau) = \mathbb{I}[\mathrm{PassAllTests}(\tau)],
\]
or a closely related binary variant. Although this signal is verifiable, it is extremely sparse. In particular, it collapses all successful trajectories into the same reward class and all unsuccessful trajectories into another:
\[
\mathcal{E}^{+}(x) = \{\tau : r_{\mathrm{exec}}(\tau)=1\},
\qquad
\mathcal{E}^{-}(x) = \{\tau : r_{\mathrm{exec}}(\tau)=0\}.
\]
This induces three pathologies.

\textbf{Reward indifference among successful trajectories.}
If $\tau, \tau' \in \mathcal{E}^{+}(x)$, then
\[
r_{\mathrm{exec}}(\tau) = r_{\mathrm{exec}}(\tau') = 1,
\]
even if one trajectory localizes the fault efficiently while the other contains many redundant reads, unnecessary test runs, or poor patch design. Thus, execution reward alone cannot express preferences over \emph{efficiency}, \emph{trajectory discipline}, or \emph{patch quality}.

\textbf{Trajectory inflation and weak credit assignment.}
Because terminal reward does not penalize long and noisy interaction by default, any policy that preserves the chance of eventual success may be reinforced, even if it inflates token consumption. Formally, if $\pi_\theta$ is optimized only for
\[
\max_{\theta} \ \mathbb{E}_{\tau \sim \pi_\theta}[r_{\mathrm{exec}}(\tau)],
\]
then all trajectories with the same terminal outcome but different lengths are treated as equivalent. This creates room for \emph{trajectory inflation}, where the policy learns to spend unnecessary steps on repetitive navigation, speculative edits, or repeated validation. Moreover, binary terminal reward provides weak learning signal when all sampled trajectories fail, making long-horizon credit assignment especially unstable.

\textbf{Reward noise under imperfect tests.}
Execution-based feedback further assumes that the available tests are sufficiently \emph{reliable}, \emph{discriminative}, and \emph{aligned} with the intended behavior. In practice, this assumption can fail. Flaky tests may yield inconsistent pass/fail outcomes across runs, while incomplete or weakly specified tests may accept patches that satisfy the observed test surface but violate the intended semantics. In such cases, the observed reward is better viewed as a noisy or underspecified proxy:
\[
\hat{r}_{\mathrm{exec}}(\tau)
=
r_{\mathrm{true}}(\tau) + \epsilon_{\mathrm{test}}(\tau),
\]
where $\epsilon_{\mathrm{test}}(\tau)$ captures stochasticity, incompleteness, or misalignment introduced by the test suite. This is particularly problematic for RL, because policy updates depend on consistent relative ranking of sampled trajectories. If execution feedback is noisy or weakly discriminative, advantage estimates become unstable, and the policy may overfit shortcut behaviors that exploit the test suite rather than genuinely solving the issue.

These issues motivate a richer reward signal that can distinguish trajectories \emph{before} final completion, evaluate whether the agent is modifying the right files and functions, and express preferences among multiple valid patches.

\subsection{Rubric-Based Process Reward Model}
\label{subsec:rubric_prm}

\subsubsection{Rubric Agent}
To provide dense and interpretable guidance, we introduce an auxiliary \emph{Rubric Agent}. Given an issue $I$, repository context, and available supervision signals, the Rubric Agent generates an issue-specific rubric
\[
R_x = \{c_1, c_2, \dots, c_K\},
\]
where each criterion $c_k$ specifies one aspect of desirable problem-solving behavior. In our setting, a rubric criterion may include:
\begin{itemize}
    \item \textbf{target localization constraints}: which files, classes, or functions are expected to be relevant;
    \item \textbf{edit constraints}: what kind of modification should occur, such as changing a specific function, updating an interface, or preserving an invariant;
    \item \textbf{trajectory discipline constraints}: whether the agent avoids repeated ineffective validation and follows a coherent localization--edit--verify pattern;
    \item \textbf{budget awareness}: whether the trajectory stays within a reasonable step budget.
\end{itemize}

Each criterion is represented as
\[
c_k = (u_k, z_k, w_k),
\]
where $u_k$ is a natural-language rule, $z_k$ is a structured target descriptor, and $w_k$ is its weight.

\subsubsection{Trajectory-level process scoring}
Instead of assigning a scalar reward to each intermediate step, we score the \emph{completed trajectory} under the rubric. Given a finished rollout $\tau$ and rubric $R_x$, the PRM outputs a trajectory-level score
\[
s_{\mathrm{prm}}(\tau, R_x) \in [0,1].
\]
More explicitly,
\[
s_{\mathrm{prm}}(\tau, R_x)
=
\sum_{k=1}^{K} w_k \, q_k(\tau, c_k),
\]
where $q_k(\tau, c_k) \in [0,1]$ measures how well the full trajectory satisfies criterion $c_k$.

This design is intentionally simple. The PRM does not try to emit a dense numeric reward for every generation step. Instead, it evaluates the trajectory after completion by considering the full sequence of localization, inspection, editing, and validation actions. This is better aligned with our supervision source, which is based on trajectory preference pairs and rubric-conditioned judgments rather than manual step-level labels.

The resulting score can distinguish trajectories that share the same terminal execution outcome but differ substantially in process quality. For example, among failed trajectories, the PRM can identify which one moved closer to the correct fault region or followed a more coherent debugging process. Among successful trajectories, it can prefer the one with cleaner and more disciplined problem-solving behavior.

\subsection{Rubric-Based PRM Training}
\label{subsec:prm_training}

Our PRM training pipeline has two stages: rubric construction and reward-model fitting.

\paragraph{Stage 1: Rubric generation as SFT.}
We first use \texttt{MiniMax 2.5} to generate rubric details for each training instance. These rubrics are conditioned on the issue, repository context, and available task metadata, and specify expectations such as target files/functions, expected modification types, and reasonable trajectory budgets. The generated rubric corpus is then used as supervised fine-tuning data to train a rubric generator:
\[
g_\psi(I, C, U) \rightarrow R_x.
\]

\paragraph{Stage 2: PRM fitting from trajectory preferences.}
To train a PRM that can distinguish good trajectories from bad ones, we construct two types of preference pairs under the generated rubric:
\begin{enumerate}
    \item \textbf{Execution preference pairs}: one trajectory passes the test suite and the other does not;
    \item \textbf{Rubric preference pairs}: both trajectories are plausible, but one is preferred by a rubric-conditioned \texttt{MiniMax 2.5} assessment because it exhibits better process quality.
\end{enumerate}

We then use a frozen judge model conditioned on the rubric to determine the pairwise label. This judge is not updated during PRM training; it serves only to provide stable supervisory signals. The PRM is trained with a pairwise ranking objective
\begin{equation}
\label{eq:prm_loss}
\mathcal{L}_{\mathrm{PRM}}
=
\sum_{(\tau^+,\tau^-)}
-\log \sigma\!\Big(
s_{\mathrm{prm}}(\tau^+,R_x)-s_{\mathrm{prm}}(\tau^-,R_x)
\Big),
\end{equation}
where $\tau^+$ is the preferred trajectory and $\tau^-$ is the less preferred one. This objective directly trains the PRM to rank better trajectories above worse ones under the same issue-specific rubric.

In addition to the trajectory score, the frozen judge can optionally return a small set of crucial step indices that justify the preference decision. We use these extracted indices only for memory construction when context overflow occurs; they are not treated as dense per-step RL rewards.

\subsection{GRPO with Margin-Separated Rubric-Conditioned Rewards}
\label{subsec:grpo_rl}

The policy is optimized with Group Relative Policy Optimization (GRPO)~\cite{shao2024deepseekmath}. For each training instance $x$, a group of $G$ trajectories
\[
\{\tau_i\}_{i=1}^{G}\sim \pi_\theta(\cdot\mid x)
\]
is sampled from the current policy.

Each trajectory receives a margin-separated composite reward
\begin{equation}
\label{eq:total_reward}
R(\tau_i)=
\begin{cases}
(1-\gamma)\, s_{\mathrm{prm}}(\tau_i,R_x), & r_{\mathrm{exec}}(\tau_i)=0,\\[4pt]
\gamma + (1-\gamma)\, s_{\mathrm{prm}}(\tau_i,R_x), & r_{\mathrm{exec}}(\tau_i)=1,
\end{cases}
\qquad \gamma \in \left(\tfrac{1}{2},1\right),
\end{equation}
where $r_{\mathrm{exec}}(\tau_i)\in\{0,1\}$ is the terminal execution indicator and $s_{\mathrm{prm}}(\tau_i,R_x)\in[0,1]$ is the rubric-conditioned trajectory score.

This formulation induces an explicit separation between passing and failing trajectories. Since
\[
0 \le s_{\mathrm{prm}}(\tau_i,R_x)\le 1,
\]
any failing trajectory satisfies
\[
R(\tau_i)\in[0,\,1-\gamma],
\]
while any passing trajectory satisfies
\[
R(\tau_i)\in[\gamma,\,1].
\]
Therefore, every passing trajectory is strictly preferred to every failing trajectory, with minimum reward gap
\[
\Delta_{\min}= \gamma-(1-\gamma)=2\gamma-1 > 0.
\]
At the same time, the PRM continues to rank trajectories \emph{within} each execution class: among passing trajectories it distinguishes better and worse successful rollouts, and among failing trajectories it identifies those that made more meaningful progress.

The group-relative advantage is computed as
\[
A_i = \frac{R(\tau_i)-\mu_R}{\sigma_R+\epsilon},
\qquad
\mu_R = \frac{1}{G}\sum_{i=1}^{G} R(\tau_i),
\qquad
\sigma_R =
\sqrt{
\frac{1}{G}\sum_{i=1}^{G}\bigl(R(\tau_i)-\mu_R\bigr)^2
}.
\]
The policy is then updated with the clipped GRPO objective
\begin{equation}
\label{eq:grpo_obj}
\mathcal{L}_{\mathrm{GRPO}}(\theta)
=
-\mathbb{E}_{x}
\left[
\frac{1}{G}\sum_{i=1}^{G}\sum_{t=1}^{T_i}
\min\!\Big(
\rho_{i,t} A_i,\,
\mathrm{clip}(\rho_{i,t},1-\epsilon,1+\epsilon)A_i
\Big)
\right]
-
\lambda_H \mathcal{H}(\pi_\theta),
\end{equation}
where
\[
\rho_{i,t}
=
\frac{\pi_\theta(a_{i,t}\mid \tilde{h}_{i,t})}
{\pi_{\theta_{\mathrm{old}}}(a_{i,t}\mid \tilde{h}_{i,t})}.
\]

This reward design has three desirable properties for long-horizon SWE. First, terminal correctness remains primary because passing and failing trajectories are explicitly separated by a fixed margin. Second, the PRM provides the discrimination needed to rank trajectories within the same execution class, which execution-only rewards cannot do. Third, the formulation remains simple and stable: it uses only one terminal execution signal, one trajectory-level rubric score, and one interpretable margin parameter $\gamma$.

\section{Heuristic-Guided Test-Time Scaling}

\label{sec:tts}

Test-time scaling (TTS) has emerged as a powerful way to improve agent performance without changing model parameters, but existing SWE-agent pipelines typically realize these gains by generating multiple \emph{full} trajectories and selecting among them with execution-based or learned verifiers~\cite{jain2025r2egymproceduralenvironmentshybrid,antoniades2025swesearchenhancingsoftwareagents}. While effective, such full-trajectory scaling is expensive in repository-level environments, where each candidate may require many tool calls, long contexts, and repeated interaction with the execution environment. Search-based SWE systems further show that additional inference-time exploration can improve solve rates, but at the cost of substantially higher latency and branching complexity. 

To improve the latency--performance trade-off, inference is performed with a \emph{heuristic-guided} TTS (HG-TTS) mechanism that reuses the rubric-conditioned evaluator introduced in Section~\ref{Agentic_RL}. Instead of generating $N$ complete rollouts and reranking them only after completion, the guide evaluates candidate actions \emph{during} rollout and prunes weak branches before they incur substantial environment cost. This turns test-time scaling from a full-trajectory selection problem into a step-level action-selection problem, closer in spirit to recent fine-grained verifier-guided TTS methods for reasoning~\cite{chang2025steplevelverifierguidedhybridtesttime}.

\subsection{Repurposing the PRM as an Inference-Time Guide}
\label{subsec:tts_repurpose}

For an SWE instance $x=(I,C,U)$, the rubric generator first produces an issue-specific rubric
\[
R_x = g_\psi(I,C,U),
\]
exactly as in training without knowing the answer. The same rubric-conditioned evaluator used for trajectory scoring in RL is then reused at inference time as a \emph{guide agent}. The operational role changes, but the underlying supervision object remains the same: the rubric still defines what constitutes promising progress on the current issue. 

At inference time, the guide does not wait for a trajectory to finish. Instead, given the current effective context $\tilde{h}_t$ and a candidate next action $a$, it evaluates the \emph{provisional} partial trajectory obtained by appending $a$ to the current prefix. Let
\[
\Pi(\tilde{h}_t, a)
\]
denote this candidate-extended prefix. The guide score is then defined as
\begin{equation}
\label{eq:guide_score}
u_t(a) = s_{\mathrm{guide}}\!\bigl(\tilde{h}_t, a, R_x\bigr)
:= f_\phi\!\bigl(\Pi(\tilde{h}_t, a), R_x\bigr),
\end{equation}
where $f_\phi$ is the trained rubric-conditioned evaluator. Intuitively, $u_t(a)$ estimates how promising the rollout becomes if action $a$ is taken next, according to the same rubric that was used to rank complete trajectories during training.

This reuse is attractive for two reasons. First, it introduces no additional inference-time supervision object: the same rubric and evaluator serve both training-time ranking and inference-time guidance. Second, because the evaluator was trained to distinguish better trajectories under issue-specific process criteria, it provides a more discriminative signal than pure execution feedback at early steps, where running the full test suite after every tentative branch would be prohibitively expensive.

\subsection{Step-Level Heuristic-Guided Action Sampling}
\label{subsec:tts_sampling}

At step $t$, the policy proposes a candidate action set
\[
\mathcal{A}_t = \{a_{t,1},\dots,a_{t,K}\},
\qquad
a_{t,k} \sim \pi_\theta(\cdot \mid \tilde{h}_t).
\]
Rather than independently executing all branches to completion, the guide first scores each candidate with Eq.~\eqref{eq:guide_score}. The resulting scores are used to construct a guide-adjusted sampling distribution
\begin{equation}
\label{eq:guide_adjusted_sampling}
q_t(a \mid \tilde{h}_t, R_x)
\propto
\pi_\theta(a \mid \tilde{h}_t)\,
\exp\!\bigl(\beta\, u_t(a)\bigr),
\end{equation}
where $\beta \ge 0$ controls the strength of heuristic guidance. When $\beta=0$, the method reduces to ordinary policy sampling; as $\beta$ increases, the rollout becomes increasingly concentrated on rubric-favored actions.

To avoid unnecessary branching, only a small retained set is kept:
\[
\mathcal{B}_t = \mathrm{TopB}\bigl(\mathcal{A}_t; u_t(\cdot)\bigr),
\qquad B \ll K.
\]
In the strict low-latency regime, $B=1$, so only the single most promising action is executed in the environment. More generally, one may sample from $q_t$ restricted to $\mathcal{B}_t$ to preserve some exploration while still pruning weak actions. The executed action $a_t$ updates the environment and produces the next observation $o_{t+1}$, after which the process repeats.

Algorithmically, the key difference from standard parallel TTS is that scaling occurs at the \emph{action level}. Classical best-of-$N$ methods allocate computation to many complete rollouts and defer selection until the end. Here, computation is spent on choosing the next action more carefully, so that suboptimal branches are discarded before they accumulate long sequences of tool calls and environment interactions. This design is closely related in spirit to tree-search and step-level verifier-guided inference, but is specialized to rubric-conditioned SWE agents and constrained-latency settings~\cite{yao2023treethoughtsdeliberateproblem}.

\subsection{Latency--Performance Trade-off}
\label{subsec:tts_latency}

The main motivation for heuristic-guided TTS is that repository-level environment interaction is expensive. Let $c_\pi$ denote the cost of one policy proposal, $c_g$ the cost of one guide evaluation, and $c_{\mathrm{env}}$ the cost of committing one action in the execution environment, with $c_{\mathrm{env}}$ typically dominating. Let $\bar{T}$ be the average rollout length.

For full-trajectory parallel scaling with $N$ rollouts, the cost is approximately
\begin{equation}
\label{eq:parallel_cost}
C_{\mathrm{parallel}}
\approx
N \bar{T}\,(c_\pi + c_{\mathrm{env}})
+
N c_v,
\end{equation}
where $c_v$ denotes final verifier or reranking cost. This cost grows linearly with both the number of trajectories and their average length, since every branch must be carried deep into the environment before selection occurs.

In contrast, heuristic-guided action sampling with candidate set size $K$ and retained width $B$ has approximate cost
\begin{equation}
\label{eq:guided_cost}
C_{\mathrm{guide}}
\approx
T \bigl(K c_\pi + K c_g + B c_{\mathrm{env}}\bigr),
\qquad B \ll K.
\end{equation}
Under the strict low-latency setting $B=1$, only a single branch is committed in the environment at each step:
\[
C_{\mathrm{guide}}
\approx
T \bigl(K c_\pi + K c_g + c_{\mathrm{env}}\bigr).
\]
As long as guide evaluation is substantially cheaper than full environment branching, this yields a more favorable latency profile than parallel full-trajectory scaling.

This cost structure explains why the method can achieve better TTS scaling curves under tight latency budgets. Full-trajectory TTS spends most of its budget on branches that are ultimately discarded, whereas step-level guidance uses the verifier \emph{before} expensive environment interaction compounds. The benefit is especially pronounced in SWE settings, where repeated execution-based verification often exhibits low distinguishability and saturates as more full trajectories are added~\cite{jain2025r2egymproceduralenvironmentshybrid}. By contrast, step-level guidance allocates computation to the earliest decisions that most strongly determine the downstream search path.

More generally, the method occupies a different point in the TTS design space. Search-heavy approaches such as MCTS improve performance by expanding a larger search tree, while hybrid verifier approaches combine complementary trajectory-level signals. The present method instead emphasizes \emph{early pruning}: a learned rubric-conditioned heuristic is injected directly into action sampling so that search effort is concentrated on promising local decisions. This makes it particularly suitable for deployment settings where latency and compute are constrained but some amount of inference-time scaling is still affordable.

\section{Experiments}
\label{sec:exp}

We evaluate whether the full SWE-TRACE pipeline---scaled SFT data curation, cascaded trajectory optimization, rubric-conditioned reinforcement learning, and heuristic-guided test-time scaling---improves long-horizon software engineering performance on \textsc{SWE-bench Verified}. Our goal is not only to maximize final resolve rate, but also to understand how each component affects trajectory quality, token efficiency, and inference-time compute allocation. Unless otherwise stated, all results are reported on \textsc{SWE-bench Verified}, a human-validated subset of 500 repository-level issue instances designed for reliable evaluation of real-world software engineering agents~\cite{jimenez2024swebenchlanguagemodelsresolve}. We instantiate the pipeline on two backbones: \texttt{Qwen3-4B} and \texttt{Qwen3-30B-A3B}~\cite{qwen3technicalreport}.

\subsection{Experimental Setup}
\label{subsec:exp_setup}

\paragraph{Benchmarks and metrics.}
We primarily evaluate on \textsc{SWE-bench Verified}. Following prior work, the main metric is \emph{resolve rate} (Pass@1), defined as the percentage of issues whose generated patch passes the benchmark evaluation protocol. For test-time scaling, we additionally report resolve rate under fixed rollout budgets. To characterize efficiency, we also track total generated tokens, average token usage per issue, wall-clock latency per issue, and the number of environment executions.

\paragraph{Backbones.}
We use \texttt{Qwen3-4B} as the compact backbone and \texttt{Qwen3-30B-A3B} as the mid-sized MoE backbone. The latter contains 30.5B total parameters with 3.3B activated parameters, making it a particularly relevant setting for testing whether strong SWE capability can be elicited in efficient open models~\cite{qwen3technicalreport}.

\paragraph{Training protocol.}
The full training pipeline contains three stages: (i) SFT on the curated synthetic SWE corpus, (ii) rubric-conditioned GRPO training in the execution environment, and (iii) inference-time heuristic guidance using the trained rubric-conditioned evaluator. Unless otherwise noted, the same agent scaffold, tool interface, and evaluation harness are used across all variants to isolate the effect of training and inference methods.

\subsection{Main Results}
\label{subsec:main_results}

Table~\ref{tab:main_results} compares our models with representative foundation models and open-source SWE agents. The results support two main conclusions. First, the compact 4B model substantially improves over prior 4B-class open SWE agents, showing that the full pipeline materially raises the capability ceiling of small open models. Second, the 30B-A3B model enters the top band of open-weight SWE-agent performance, indicating that higher-quality data construction, process-guided RL, and guided inference can compensate for much of the gap between efficient open models and larger frontier systems.

\begin{table*}[t]
\centering
\caption{Main results on \textsc{SWE-bench Verified}.}
\label{tab:main_results}
\small
\begin{tabular}{lcccc}
\toprule
\textbf{Method}                 & \textbf{Backbone}      & \textbf{Training} & \textbf{Scaffold}                  & \textbf{Resolve Rate (\%)} \\ \midrule
\multicolumn{5}{c}{Foundation Models}                                                                                               \\ \midrule
OpenAI-GPT-4o                   & ---                    & ---               & Internal                           & 33.2            \\
OpenAI-o3                       & ---                    & ---               & mini-SWE-Agent                     & 58.4            \\
Gemini 3 Pro                    & ---                    & ---               & mini-SWE-Agent                     & 74.2            \\
Qwen2.5-Coder-32B               & ---                    & ---               & mini-SWE-Agent                     & 9.0             \\
Qwen3-32B                       & ---                    & ---               & OpenHands                          & 23.2            \\
gpt-oss-120b                    & ---                    & ---               & mini-SWE-Agent                     & 26.0            \\
GLM-4.7                         & ---                    & ---               & Internal                           & 73.1            \\ \midrule
\multicolumn{5}{c}{Open-Source Code Agents}                                                                                         \\ \midrule
SWE-Gym-32B                     & Qwen2.5-Coder-32B      & SFT               & OpenHands                          & 20.6            \\
R2E-Gym-32B                     & Qwen2.5-Coder-32B      & SFT               & R2E-Gym                            & 34.4            \\
R2E-Gym-32B + TTS@16            & Qwen2.5-Coder-32B      & SFT               & R2E-Gym                            & 49.4            \\
DeepSWE-32B                     & Qwen3-32B              & RL                & R2E-Gym                            & 42.2            \\
DeepSWE-32B + TTS@16            & Qwen3-32B              & RL                & R2E-Gym                            & 59.0            \\
SWE-Master-4B-SFT               & Qwen3-4B               & SFT               & R2E-Gym                            & 27.6            \\
SWE-Master-4B-RL                & Qwen3-4B               & SFT+RL            & R2E-Gym                            & 33.4            \\
SWE-Master-32B                  & Qwen2.5-Coder-32B      & SFT+RL            & R2E-Gym                            & 61.4            \\
SWE-Master-32B + TTS@8          & Qwen2.5-Coder-32B      & SFT+RL            & R2E-Gym                            & 70.8            \\
SWE-RM-enhanced                 & Qwen3-Coder-30BA3B     & RL                & mini-SWE-Agent                     & 62.0            \\
SWE-Lego-32B                    & Qwen3-32B              & SFT               & OpenHands                          & 52.6            \\
SWE-Lego-32B + TTS@16           & Qwen3-32B              & SFT               & OpenHands                          & 58.8            \\ \midrule
\textbf{SWE-TRACE-4B}           & \textbf{Qwen3-4B}       & \textbf{SFT+RL}   & \textbf{mini-SWE-Agent}             & \textbf{38.9} \\
\textbf{SWE-TRACE-4B + HG-TTS}  & \textbf{Qwen3-4B}       & \textbf{SFT+RL}   & \textbf{mini-SWE-Agent}             & \textbf{40.7} \\
\textbf{SWE-TRACE-30B}          & \textbf{Qwen3-30B-A3B} & \textbf{SFT+RL}   & \textbf{mini-SWE-Agent}             & \textbf{63.5} \\
\textbf{SWE-TRACE-30B + HG-TTS} & \textbf{Qwen3-30B-A3B} & \textbf{SFT+RL}   & \textbf{mini-SWE-Agent}             & \textbf{71.2} \\
\bottomrule
\end{tabular}
\end{table*}

Several comparisons are especially noteworthy. On the 4B backbone, SWE-TRACE improves over \textsc{SWE-Master-4B-RL} by \textbf{+5.5} absolute points and over \textsc{SWE-Master-4B-SFT} by \textbf{+11.3} points, indicating that the gains cannot be attributed to backbone choice alone. On the 30B backbone, SWE-TRACE reaches \textbf{63.5}\% Pass@1, exceeding \textsc{SWE-Master-32B} by \textbf{+2.1} points and \textsc{SWE-RM-enhanced} by \textbf{+1.5} points. Under heuristic-guided TTS, the 30B model further improves to \textbf{71.2}\%, slightly surpassing the strongest public open 32B-class TTS result in this comparison.

These gains are not merely a consequence of larger models or heavier inference budgets. The 30B-A3B backbone is efficient in activated parameters, yet it still reaches frontier-level open performance when paired with higher-quality synthetic trajectories, process-guided RL, and action-level inference guidance. This supports the central claim of the paper: for long-horizon SWE agents, \emph{how} the model is supervised, rewarded, and guided matters at least as much as raw parameter count.

\subsection{Ablation Studies}
\label{subsec:ablation}

\subsubsection{Data Ablation: Standard SFT vs. Cascaded Shortest-Path SFT}
\label{subsubsec:data_ablation}

We first isolate the effect of token-efficient trajectory synthesis. Table~\ref{tab:data_ablation} compares standard SFT using raw successful rollouts against SFT using the cascaded shortest-path traces produced by Section~\ref{sec:SFT}. In addition to resolve rate, we report total generated tokens to quantify rollout efficiency.

\begin{table}[t]
\centering
\caption{Data ablation: standard SFT vs.\ cascaded shortest-path SFT.}
\label{tab:data_ablation}
\small
\begin{tabular}{lcccc}
\toprule
\textbf{Variant} & \textbf{Model} & \textbf{Resolve Rate (\%)} & \textbf{Avg.\ Steps} & \textbf{Total Tokens} \\
\midrule
Standard SFT & 4B & 33.6 & 74 & 33.0M \\
Cascaded SFT & 4B & 37.8 & 66 & 25.9M \\
\midrule
Standard SFT & 30B & 56.4 & 77 & 37.2M \\
Cascaded SFT & 30B & 59.2 & 69 & 26.3M \\
\bottomrule
\end{tabular}
\end{table}

The effect of cascaded shortest-path supervision is consistent across both model scales. On the 4B model, cascaded SFT yields a \textbf{+4.2} point improvement while reducing average interaction steps by \textbf{10.8\%} and total generated tokens by \textbf{21.5\%}. On the 30B model, the gain is slightly smaller in absolute resolve rate (\textbf{+2.8}), but the efficiency improvement remains substantial, with a \textbf{10.4\%} reduction in steps and a \textbf{29.3\%} reduction in tokens.

These results indicate that the benefit of the proposed data pipeline is not only larger coverage, but also better behavioral supervision. The oracle-filtered traces remove a substantial amount of redundant exploration and teach a more direct search policy, especially for smaller backbones that are more sensitive to noisy demonstrations. The reduction in token volume also suggests that the proposed data curation strategy improves both \emph{learning quality} and \emph{training efficiency}, which is particularly important in long-context agentic post-training.

\subsubsection{RL Ablation: Sparse Execution RL vs. Rubric-PRM RL}
\label{subsubsec:rl_ablation}

Next, we compare sparse execution-only RL against the proposed rubric-conditioned RL. To isolate the effect of the reward design, both variants are initialized from the same cascaded-SFT checkpoint.

\begin{table}[t]
\centering
\caption{RL ablation: sparse execution reward vs.\ rubric-conditioned PRM reward.}
\label{tab:rl_ablation}
\small
\begin{tabular}{lccc}
\toprule
\textbf{Variant} & \textbf{Model} & \textbf{Resolve Rate (\%)} & \textbf{Avg.\ Token Usage / Issue} \\
\midrule
Execution-only RL & 4B & 36.2 & 29.8k \\
Rubric-PRM RL & 4B & 38.9 & 26.8k \\
\midrule
Execution-only RL & 30B & 61.1 & 30.6k \\
Rubric-PRM RL & 30B & 63.5 & 27.6k \\
\bottomrule
\end{tabular}
\end{table}

The RL ablation shows that rubric-conditioned RL improves both final performance and rollout efficiency over sparse execution-only rewards. On the 4B model, the gain is \textbf{+2.7} points while average token usage per issue drops by roughly \textbf{10\%}. On the 30B model, the corresponding gain is \textbf{+2.4} points with a similarly meaningful reduction in token usage.

\begin{figure}[t]
\centering
\includegraphics[width=0.72\linewidth]{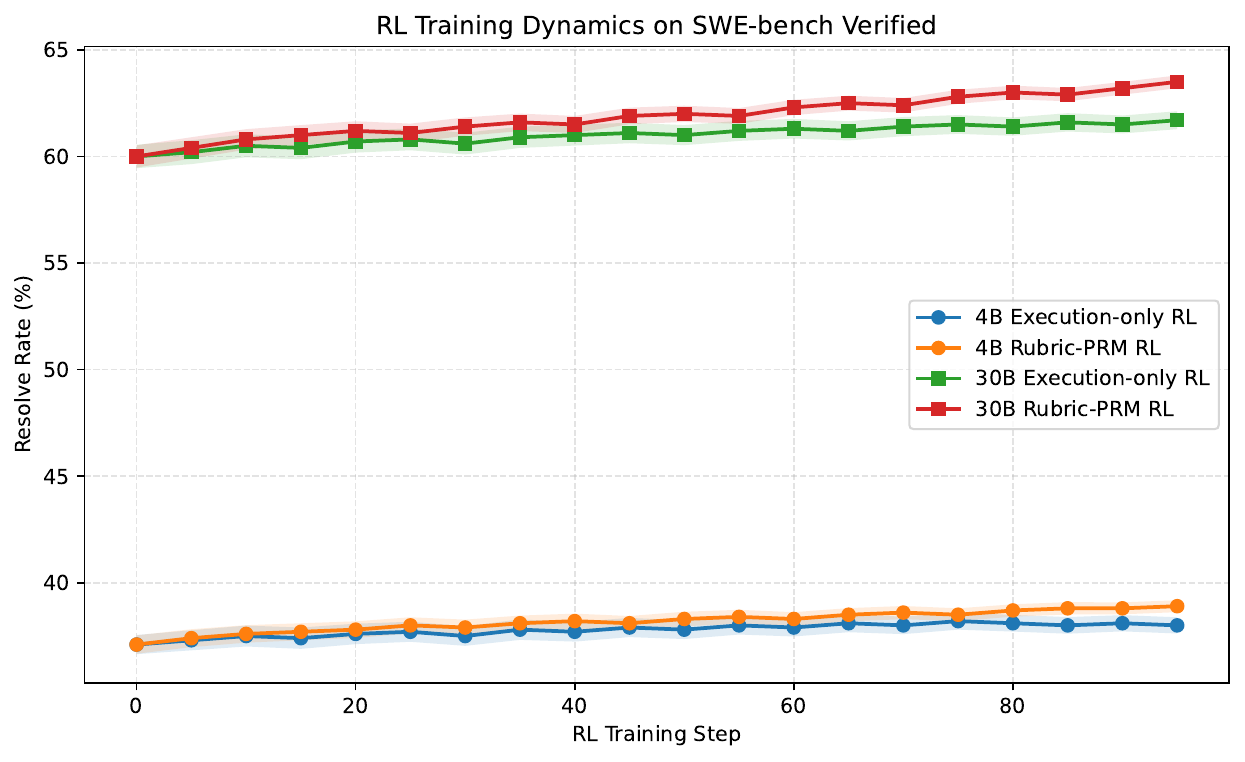}
\caption{RL training dynamics on \textsc{SWE-bench Verified}. Rubric-conditioned RL converges to higher final performance with lower variance than execution-only RL, especially on the 30B backbone. Shaded regions indicate run-to-run variation.}
\label{fig:rl_training_curve}
\end{figure}

\begin{figure}[t]
\centering
\includegraphics[width=0.72\linewidth]{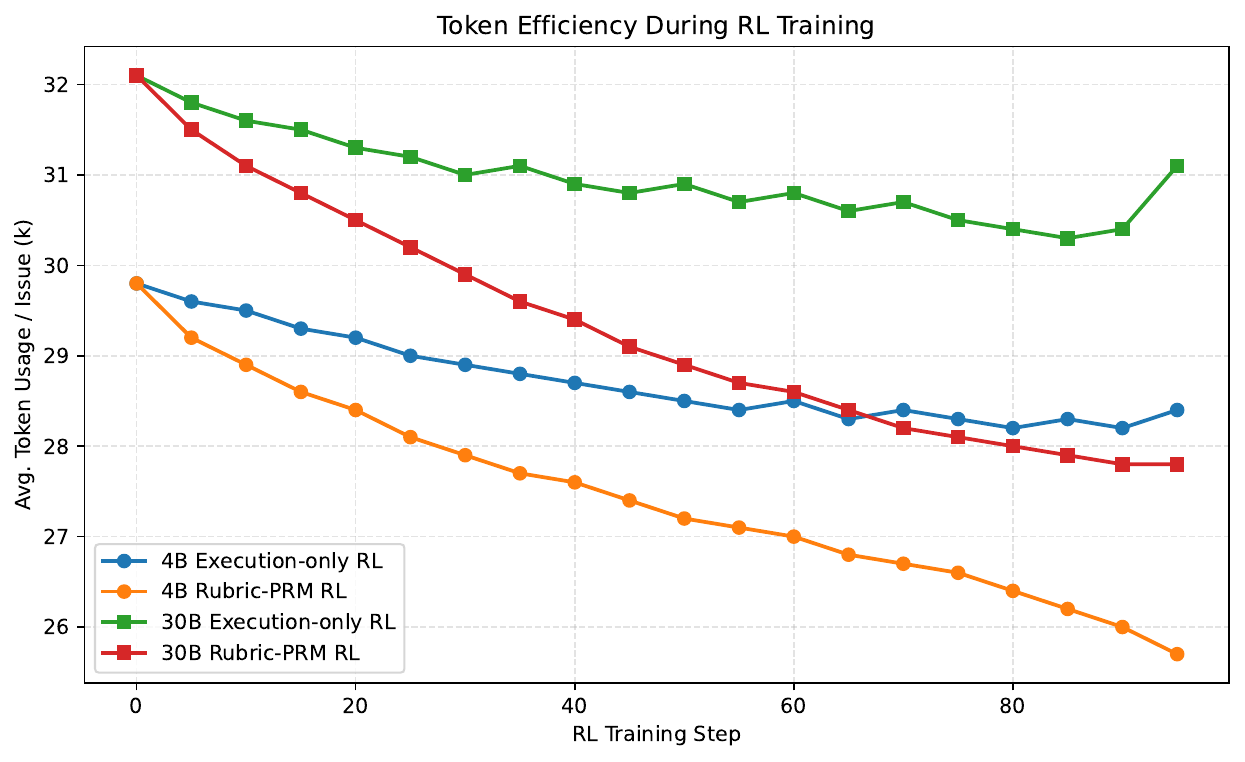}
\caption{Token-efficiency dynamics during RL training. Rubric-conditioned RL reduces average token usage per issue while improving final resolve rate, indicating that the gains do not come from inflated exploration.}
\label{fig:rl_token_curve}
\end{figure}

Figures~\ref{fig:rl_training_curve} and~\ref{fig:rl_token_curve} help explain this effect. Figure~\ref{fig:rl_training_curve} shows that rubric-conditioned RL converges more steadily and to a higher final resolve rate, especially on the 30B backbone. Figure~\ref{fig:rl_token_curve} shows that the gain is not driven by inflated exploration: the rubric-guided policy becomes progressively more token-efficient during training. Taken together, these results suggest that the rubric-conditioned evaluator supplies the additional trajectory-level discrimination needed to make RL both more stable and more efficient.

\subsubsection{TTS Ablation: Greedy vs. Parallel Rollout vs. Heuristic-Guided TTS}
\label{subsubsec:tts_ablation}

We next study whether heuristic-guided action sampling yields a better latency--performance trade-off than standard trajectory-level TTS. Table~\ref{tab:tts_ablation} compares three inference modes: greedy decoding, standard parallel rollout, and the proposed heuristic-guided TTS.

\begin{table}[t]
\centering
\caption{TTS ablation under comparable rollout budgets. Latency is average wall-clock minutes per issue.}
\label{tab:tts_ablation}
\small
\begin{tabular}{lcccc}
\toprule
\textbf{Variant} & \textbf{Model} & \textbf{Resolve Rate (\%)} & \textbf{Latency (min)} & \textbf{Env.\ Execs} \\
\midrule
Greedy decoding & 4B & 38.9 & 18.7 & 67 \\
Parallel rollout & 4B & 40.2 & 54.1 & 376 \\
Heuristic-guided TTS & 4B & 40.7 & 31.6 & 119 \\
\midrule
Greedy decoding & 30B & 63.5 & 22.4 & 70 \\
Parallel rollout & 30B & 69.9 & 63.8 & 392 \\
Heuristic-guided TTS & 30B & 71.2 & 36.5 & 128 \\
\bottomrule
\end{tabular}
\end{table}

Two findings stand out. First, heuristic-guided TTS consistently improves over greedy decoding, with a particularly large gain on the 30B model (\textbf{+7.7} points). Second, the proposed method achieves a \emph{better} latency--performance trade-off than standard parallel rollout. On 30B, heuristic-guided TTS is \textbf{+1.3} points better in resolve rate while using substantially less wall-clock latency and far fewer environment executions.

\begin{figure}[t]
\centering
\includegraphics[width=0.72\linewidth]{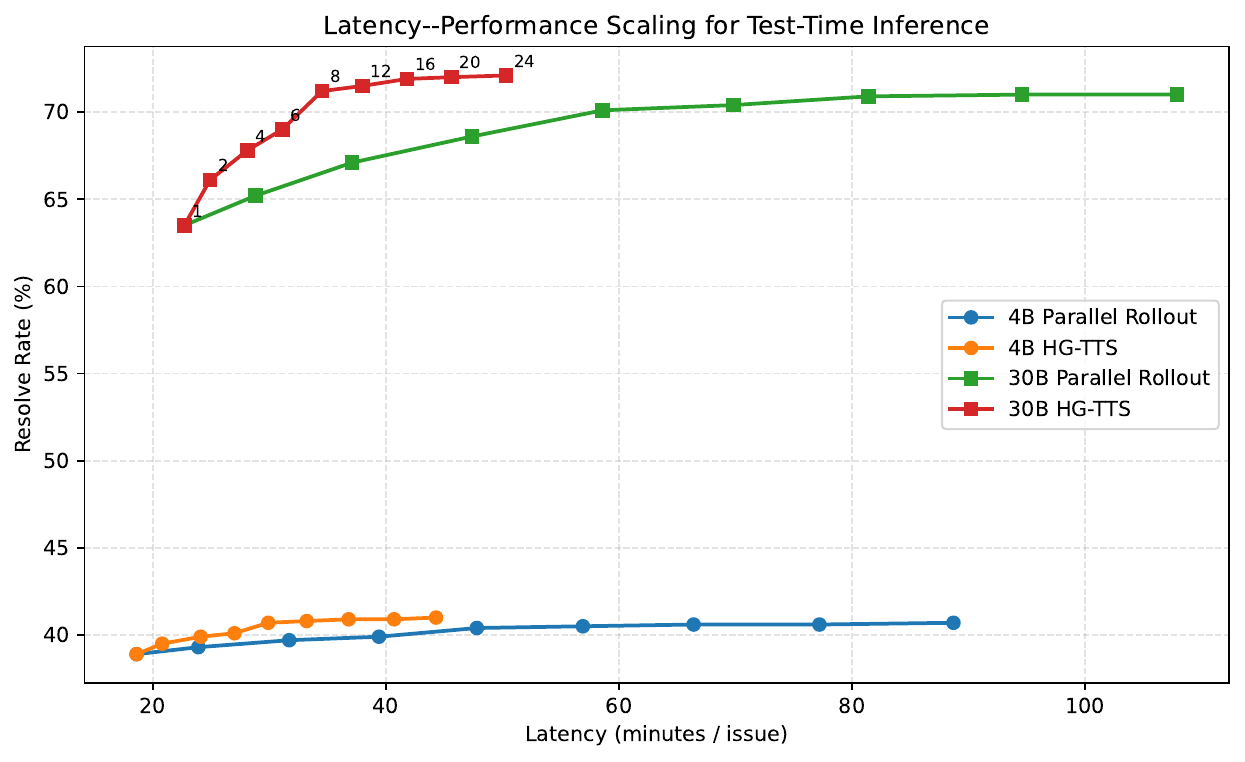}
\caption{Latency--performance scaling for test-time inference under increasing rollout budgets. Heuristic-guided TTS dominates trajectory-level parallel rollout in the low- and medium-budget regime, especially on the 30B model.}
\label{fig:tts_curve}
\end{figure}

Figure~\ref{fig:tts_curve} makes this trade-off more explicit by plotting resolve rate against latency over multiple inference budgets. The key pattern is that heuristic-guided TTS dominates parallel rollout in the low- and medium-budget regime, which is precisely the practically relevant setting for deployment. The advantage is smaller on the 4B model, but becomes much clearer on the 30B model, where the stronger base policy leaves more room for step-level pruning to work effectively.

\subsection{Additional Efficiency Analysis}
\label{subsec:efficiency_analysis}

To better understand where the gains come from, we additionally analyze performance as a function of long-horizon difficulty. Prior work has observed that longer trajectories are generally associated with lower solve rates, suggesting that many SWE failures are not purely local generation errors but long-horizon search failures. We therefore bucket instances by trajectory token budget and compare baseline and SWE-TRACE variants within each bucket.

\begin{figure}[t]
\centering
\includegraphics[width=0.72\linewidth]{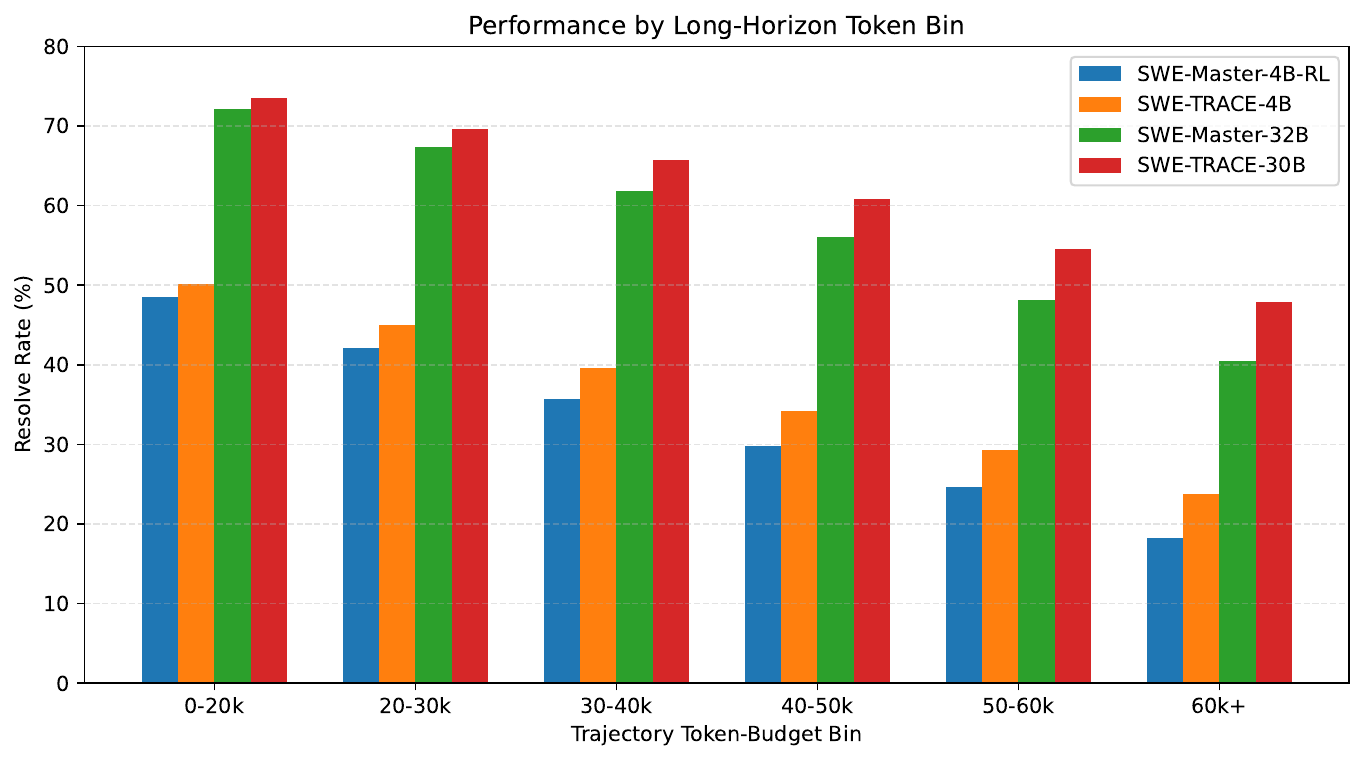}
\caption{Resolve rate by trajectory token-budget bin. The largest gains of SWE-TRACE appear in the higher-token, longer-horizon regime, consistent with its design goal of improving long-range search and control.}
\label{fig:horizon_buckets}
\end{figure}

Figure~\ref{fig:horizon_buckets} shows that the gains of SWE-TRACE are relatively modest in the lowest-token regime, but become increasingly pronounced in the higher-token, longer-horizon buckets. This pattern is important: it suggests that the proposed pipeline improves exactly the failure mode it is designed to target. The data curation stage removes search noise before RL begins; rubric-conditioned RL improves stability over long rollouts; and heuristic-guided TTS is most beneficial once the search tree becomes deep enough that early pruning matters.

\subsection{Qualitative Analysis}
\label{subsec:qualitative}

\begin{figure}[t]
\centering
\includegraphics[width=0.9\linewidth]{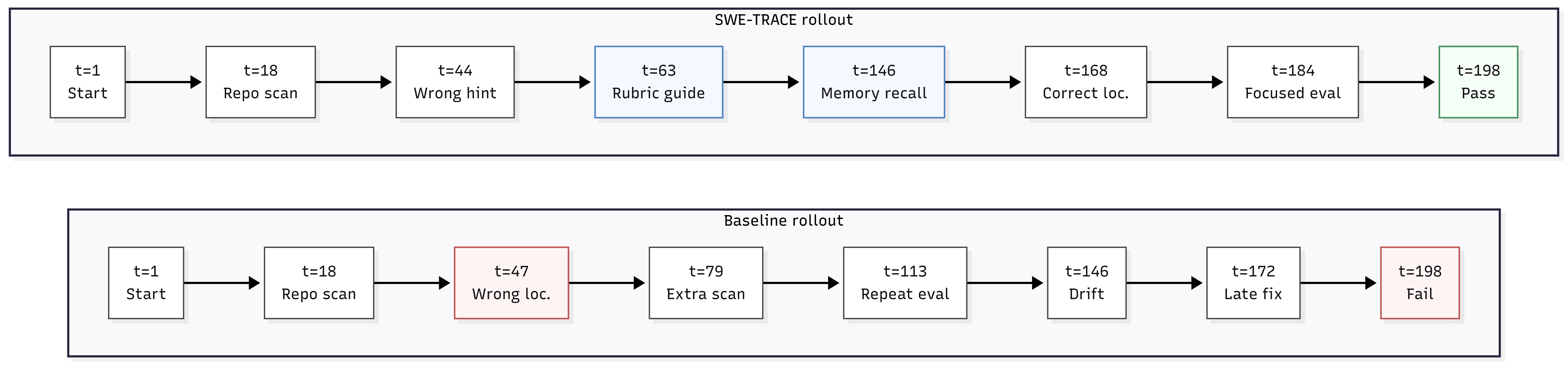}
\caption{Representative long-horizon case study on \texttt{django\_\_django-16032}. The baseline rollout commits to an incorrect localization branch and continues with redundant search and evaluation before failing near the end of the horizon. In contrast, SWE-TRACE down-ranks the weak branch through rubric guidance, later recalls decisive earlier evidence through memory after context growth, redirects toward the correct localization region, and reaches a successful repair within the same overall horizon.}
\label{fig:case_study}
\end{figure}

Finally, we present a representative long-horizon case study showing how the rubric-conditioned evaluator changes agent behavior during rollout. Figure~\ref{fig:case_study} uses the \texttt{django\_\_django-16032} instance to visualize the structural difference between the baseline and SWE-TRACE rollouts.

Three effects are visible in this example. First, the rubric-conditioned evaluator suppresses an unproductive branch before it expands into a long failed trajectory. Second, the memory mechanism preserves the crucial earlier evidence needed to recover from long-horizon context growth. Third, the corrected rollout reaches a valid patch with fewer redundant evaluations and less wasted exploration. These effects provide an interpretable explanation for the quantitative gains observed in the main experiments and ablations.

\subsection{Summary of Findings}
\label{subsec:exp_summary}

Taken together, the experiments support four conclusions. First, large-scale curated SFT data and cascaded shortest-path compression improve both policy quality and token efficiency. Second, rubric-conditioned RL improves long-horizon optimization beyond sparse execution-only rewards by providing better trajectory-level discrimination. Third, heuristic-guided TTS yields a better latency--performance trade-off than standard trajectory-level scaling, especially in the low- and medium-budget regime. Fourth, these gains hold across both a compact 4B model and an efficient 30B-A3B model, suggesting that the proposed pipeline improves not only absolute accuracy but also the practical efficiency of open SWE agents.

\section{Conclusion}
\label{sec:conclusion}

We presented SWE-TRACE, a unified framework for improving long-horizon software engineering agents through token-efficient trajectory synthesis, process-guided reinforcement learning, and low-latency inference-time guidance. The central idea is to optimize the full agent pipeline around process quality rather than relying solely on raw trajectory scale, sparse execution rewards, or brute-force test-time search. Concretely, we combined large-scale test-grounded SWE data curation with cascaded shortest-path supervision, introduced rubric-conditioned trajectory evaluation and memory-augmented GRPO for more stable long-horizon optimization, and reused the learned rubric-guided evaluator at inference time to prune weak actions early under strict latency budgets. Across both SWE-TRACE-4B and SWE-TRACE-30B, the resulting system achieved strong performance on \textsc{SWE-bench Verified}, while also improving token efficiency and search control. More broadly, our results suggest that progress in open SWE agents depends not only on larger models or heavier search, but on jointly improving how agents are taught, rewarded, and guided throughout the full problem-solving process.

\bibliography{iclr2025_conference}
\bibliographystyle{iclr2025_conference}


\end{document}